\documentclass[10pt]{article}
\usepackage[dvips]{graphicx}
\usepackage{amsmath}
\newcommand{\rmd}{\mathrm{d}}
\newcommand{\rmi}{\mathrm{i}}
\newcommand{\bi}[1]{\textbf{\textit{#1}}}

\begin{document}
\title{\textbf{Large radius exciton in single-walled carbon nanotubes}}
\author{Vadym M. Adamyan\footnote{E-mail: vadamyan@paco.net}, Oleksii A. Smyrnov\footnote{E-mail: smyrnov@onu.edu.ua}\\
\emph{Department of Theoretical Physics, Odessa I. I. Mechnikov
National University,}\\ \emph{2 Dvoryanskaya St., Odessa 65026,
Ukraine}}
\date{June 30, 2007}
\maketitle
\begin{abstract}
The spectrum of large radius exciton in an individual
semiconducting single-walled carbon nanotube (SWCNT) is described
within the framework of elementary potential model, in which
exciton is modeled as bound state of two oppositely charged
quasi-particles confined on the tube surface. Due to the parity of
the interaction potential the exciton states split into the odd
and even series. It is shown that for the bare and screened
Coulomb electron-hole ($e$-$h$) potentials the binding energy of
even excitons in the ground state well exceeds the energy gap. The
factors preventing the collapse of single-electron states in
isolated semiconducting SWCNTs are discussed.
\end{abstract}

PACS number(s): 78.67.Ch

\section{Introduction}
\setcounter{equation}{0}

Many experimental papers on the optical absorption in SWCNTs
describe obtained results in terms of band-to-band direct
transitions between single particle states, though it is clear
that the inherent to 1D systems strong interparticle interaction
cannot be neglected. It seems obvious that the strong
electron-hole attraction should bind electron-hole pairs in SWCNTs
into Wannier-Mott like excitons. Moreover, the exciton
contributions were already revealed experimentally in optical
absorption spectra~\cite{ichida1},~\cite{ichida2}, and in spectra
of fluorescence~\cite{fluor}-\cite{bachilo2} of individual SWCNTs,
as well the exciton properties were studied by the Raman
spectroscopy~\cite{wang}. There are also some works devoted to the
theoretical study of excitons in CNTs~\cite{ando1}-\cite{chang}.
However, as it follows from results of the latter a simple
translation of basic hydrogen-like models of 3D large radius
excitons fails to function in one dimension without a certain
specification. Remind that once the centrum of mass has been
removed and the screening effect from the tube charges is ignored
the 1D model exciton Hamiltonian may be formally given by the
expression
\begin{equation}\label{1.1}
\widehat{H}=-\frac{\hbar^2}{2\mu}\frac{\rmd^2}{\rmd
z^2}-\frac{e^2}{\varepsilon}\frac{1}{|z|},
\end{equation}
where $\mu$ is the exciton reduced effective mass and
$\varepsilon$ is the dielectric constant of medium surrounding the
tube. The functions $\phi(z)$ from the domain of differential
expression~(\ref{1.1}) in $\mathbf{L}_2(-\infty,\infty)$ are twice
differentiable on the semi-axis $(-\infty,0)$, $(0,\infty)$, and
belong together with their derivatives for each $\delta>0$ to the
subspaces $\mathbf{L}_2(-\infty,-\delta)$,
$\mathbf{L}_2(\delta,\infty)$, and also satisfy a certain boundary
conditions at the point $z=0$, which provide the self-adjointness
of $\widehat{H}$. If there are no reasons for breaking of the tube
reflection symmetry or, in other words, of the action and reaction
law, then only those self-adjoint extensions of the above
differential operator are physically admissible, for which the
subspaces of even and odd functions from
$\mathbf{L}_2(-\infty,\infty )$ are invariant with respect to
$\widehat{H}$.

For less singular than the bare Coulomb even potentials
$V(z)=V(|z|)$, for example, for potentials satisfying the
condition
\[
\int\limits_0^L|V(z)|^2\rmd z<\infty,\qquad 0<L<\infty,
\]
the corresponding Hamiltonians $\widehat{H}$ could be represented
as the sum of the standardly defined self-adjoint operator of
kinetic energy and the subordinated operator of potential energy
that is the multiplication operator by $V(z)$. In such cases the
functions from the domains $\mathcal{D}(\widehat{H})$ of
$\widehat{H}$ are continuous and have continuous first derivative
at $z=0$, and the corresponding odd and even functions from
$\mathcal{D}(\widehat{H})$ should satisfy the natural boundary
conditions $\phi(0)=0$ and $\phi'(0)=0$, respectively. The spectra
of bound states are then the set of negative eigenvalues of the
boundary problem for the differential equation
\[
-\frac{\hbar^2}{2\mu}\frac{\rmd^2}{\rmd
z^2}\phi(z)-V(z)\phi(z)=E\phi(z)
\]
on the semi-axis $(0,\infty)$ with the boundary conditions
$\phi(0)=0$ for the odd series and $\phi'(0)=0$ for the even
series, respectively.

For the Coulomb potential in one dimension the operator of
potential energy is not subordinated to that of kinetic energy and
the Hamiltonian $\widehat{H}$ cannot be simply represented as
their sum, and hence the Hamiltonian in this case becomes
indeterminate. Among self-adjoint extensions of the differential
operator~(\ref{1.1}) there are infinitely many of those for which
subspaces of odd and even functions are invariant. These
extensions differ in self-adjoint boundary conditions
at $z=0$ and, accordingly,  they can be distinguished by energies of
their ground states. Since functions from the domain of any such extension
are non-differentiable at $z=0$, it is impossible without
additional physical considerations to single out the unique
"correct" among suitable self-adjoint boundary conditions at
$z=0$.

For the bare Coulomb potential one of possible extensions
compatible with $z$-inversion symmetry is the decaying extension
$\widehat{H}_0$ defined by the boundary condition: $\phi(0)=0$,
that is for any $\phi$ from the domain of $\widehat{H}_0$ we have
\begin{equation}\label{1.2}
\left\{ \begin{array}{c}
\left(\widehat{H}_0\phi\right)(z)=-(\hbar^2/2\mu)\phi''(z)+(e^2/\varepsilon z)\phi(z),\ z<0;\\
\\
\left(\widehat{H}_0\phi\right)(z)=-(\hbar^2/2\mu)\phi''(z)-(e^2/\varepsilon z)\phi(z),\ z>0;\\
\\
\phi(-0)=\phi(+0)=0.
\end{array}\right.
\end{equation}

This Hamiltonian was obtained in~\cite{loudon},~\cite{cornean} as
a result of formal passage to the limit for some sequences of 1D
Hamiltonians with regularized at $z=0$ Coulomb potentials. Note,
that not only subspaces of odd and even functions, but also the
subspaces of functions with supports on the positive and negative
semi-axis are invariant with respect to $\widehat{H}_0$.  In other
words, for the zero boundary condition at the origin
$\widehat{H}_0$ is isomorphic to the orthogonal sum of two reduced
"classic" Schr\"{o}dinger operators for $s$-states of the hydrogen
atom. As follows, the negative spectrum for this Hamiltonian is
the Balmer series, each eigenvalue of which is doubly degenerate.
It is worth mentioning, that if the states of electron-hole pair
would be governed by the Hamiltonian $\widehat{H}_0$, then
excitons in a tube would be subdivided into conserved "left" and
"right" ones subject to the positional relationship of electron
versus hole.

In Section 2 of this paper we consider another extension
$\widehat{H}_1$, which in the odd sector coincides with
$\widehat{H}_0$ but in the even sector is defined on the subset of
continuous functions satisfying at $z=0$ the boundary condition
\begin{equation}\label{1.3}
\begin{split}
&\lim_{z\uparrow0}\frac{\rmd}{\rmd
z}\left[\left(1-2Az\ln(2A|z|)\right)\phi(z)\right]\\
&=\lim_{z\downarrow0}\frac{\rmd}{\rmd
z}\left[\left(1+2Az\ln(2Az)\right)\phi(z)\right]=0;\
A=e^2\mu/\hbar^2.
\end{split}
\end{equation}

For $\widehat{H}_1$ the spectrum of bound states of the even
series appeared to be close to that for the two-dimensional
hydrogen atom~\cite{2Dhydr} for the states with zero angular
momentum. This fact as well as the transition of~(\ref{1.3}) to
the Neumann condition $\phi'(-0)=\phi'(+0)=0$ as $A\rightarrow0$
is not yet a valid reason to consider $\widehat{H}_1$ as an
appropriate primordial Hamiltonian for the large radius exciton in
nanotubes. However, the modified electron-hole interaction
potential $V(z)$, that accounts, that these particles actually are
not pointwise and their charges are smeared along infinitesimal
narrow bands on the tube surface, appeared to be locally
quadratically integrable. In the case of nanotubes of small
diameters the Hamiltonian with this potential gives the energies
of ground and first excited states of the standardly defined even
series, which differ slightly from those for $\widehat{H}_1$.

However, it turned out that the ground state energy of even
excitons, calculated for individual semiconducting carbon
nanotubes in vacuum with this potential and without account of the
effect of screening by the nanotube electrons, are just two times
greater of the energy gaps. Therefore, in Sections 3 and 4 we
consider different forms of screening of the electron-hole
interaction inside individual semiconducting nanotubes (e.g., we
calculate the dielectric function of individual SWCNTs). Results
on the ground state of even excitons, given in Section 5 for some
individual carbon nanotubes, show that the account of screening
does not help and the binding energy of even excitons remains
greater of the energy gap. This may mean the instability of
single-electron states in isolated semiconducting carbon nanotubes
in the vicinity of the energy gap against the exciton formation.
In the last section of the paper we discuss factors preventing the
collapse of single-electron states in isolated semiconducting
SWCNTs.

\section{Exciton spectrum and eigenfunctions in the Coulomb limit}
\setcounter{equation}{0}

Let $\psi_\mathrm{v}(k,\bi{r})$ and $\psi_\mathrm{c}(k,\bi{r})$ be
the Bloch wave functions of the valence and conduction band
electrons of a semiconducting nanotube, respectively. Remind that
\[
\psi_{\mathrm{v},\mathrm{c}}(k,\bi{r})=\exp(\rmi
kz)u_{\mathrm{v},\mathrm{c}}(k,\bi{r}),\nonumber
\]
where $u_{\mathrm{v},\mathrm{c}}(k,\bi{r})$ are periodic functions
with the period $a$ along the tube axis, which is assumed to
coincide with the $z$-axis. The wave functions of rest exciton
can be represented as the following superposition:
\begin{equation}\label{2.1}
\Psi(\bi{r}_1,\bi{r}_2)=\int\limits_{-\pi/a}^{\pi/a}\Phi(k)\psi^*_\mathrm{c}(k,\bi{r}_1)\psi_\mathrm{v}(k,\bi{r}_2)\rmd
k.
\end{equation}
The envelope function $\Phi(k)$ in~(\ref{2.1})
satisfies the equation
\begin{equation}\label{2.2}
(\epsilon_\mathrm{c}(k)-\epsilon_\mathrm{v}(k))\Phi(k)+\frac{a}{2\pi}\int\limits_{-\pi/a}^{\pi/a}J(k,k')\Phi(k')\rmd
k'= E_\mathrm{{exc}}\Phi(k),
\end{equation}
where $\epsilon_\mathrm{v}(k)$ and $\epsilon_\mathrm{c}(k)$ are
band energies of electrons with quasi-momentum $k$ and the kernel
\begin{equation}
\begin{split}
J(k,k')=&-\lim_{L\rightarrow\infty}\frac{a}{L}\int\limits_{\mathrm{E}_3^L}\int\limits_{\mathrm{E}_3^L}\psi_\mathrm{c}(k,\bi{r}_1)\psi^*_\mathrm{v}(k,\bi{r}_2)
\frac{e^2}{|\bi{r}_1-\bi{r}_2|}\psi^*_\mathrm{c}(k',\bi{r}_1)\psi_\mathrm{v}(k',\bi{r}_2)\rmd\bi{r}_1\rmd\bi{r}_2,\\
&\mathrm{E}_3^L=\mathrm{E}_2\times(0<z<L),\nonumber
\end{split}
\end{equation}
corresponds to the two-particle system interacting with each other
through the bare Coulomb potential.

In the so-called long-wave approximation~(\ref{2.2}) takes the
form
\begin{equation}\label{2.3}
\left(E_\mathrm{g}+\frac{\hbar^2k^2}{2\mu}\right)\Phi(k)+\frac{1}{2\pi}\int\limits_{-\infty}^{\infty}\widetilde{V}(k-k')\Phi(k')\rmd
k'= E_{\mathrm{exc}}\Phi(k),
\end{equation}
where $E_\mathrm{g}$ and $\mu$ are the gap width and the reduced
effective mass of electron and hole, respectively, and
$\widetilde{V}(k)$ is the Fourier transform of the effective
potential
\begin{equation}\label{2.4}
\begin{split}
V(z)=-&\int\limits_{\mathrm{E}_3^a}\int\limits_{\mathrm{E}_3^a}\frac{e^2}{\left((x_1-x_2)^2+(y_1-y_2)^2+(z+z_1-z_2)^2\right)^{1/2}}\\
&\times|u_\mathrm{c}(0,\bi{r}_1)|^2|u_\mathrm{v}(0,\bi{r}_2)|^2\rmd\bi{r}_1\rmd\bi{r}_2.
\end{split}
\end{equation}
We see from~(\ref{2.3}) that~(\ref{2.2}) is equivalent to 1D
Schr\"{o}dinger equation
\begin{equation}\label{2.5}
-\frac{\hbar^{2}}{2\mu}\phi''(z)+V(z)\phi(z)=\mathcal{E}\phi(z),\
\mathcal{E}=E_\mathrm{exc}-E_\mathrm{g},
\end{equation}
on the real axis. Note, that independently on the tube radius and
chirality
\[
\left.V(z)\right|_{z\rightarrow\pm\infty}\simeq-e^2/|z|+o\left(1/|z|\right).
\]
Contrary to 3D case~\cite{land} direct using of the
equation~(\ref{2.5}) with potential $V_0(z)=-e^2/|z|$ for
modelling of exciton states in nanotubes is impossible without a
more accurate definition of the exciton Hamiltonian for short
distances between electron and hole. The matter is that due to the
Coulomb singularity of $V_0(z)$ the one-dimensional
Schr\"{o}dinger operator for a hydrogen-like system remains
indeterminate without imposing of certain (self-adjoint) boundary
conditions onto wave functions at the point $z=0$. So to define
the 1D exciton Hamiltonian we should either specify such a
boundary condition or "soften" the singularity of potential at
short distances with respect to the expression~(\ref{2.4}) and
screening effects from the tube's electrons. As it was mentioned
above, for parity of the Coulomb potential the exciton states are
split into two series: even $\phi(-z)=\phi(z)$ and odd
$\phi(-z)=-\phi(z)$. Despite of the Coulomb singularity at $z=0$,
any solution of the equation
\begin{equation}\label{2.6}
\begin{split}
&\frac{\rmd^2\phi}{\rmd z^2}+\left(\frac{2k\kappa}{|z|}-\kappa^2\right)\phi=0,\\
&k=\mu e^2/\kappa\hbar^2,\ \kappa=\sqrt{2\mu
|\mathcal{E}|/\hbar^2}
\end{split}
\end{equation}
has continuous left and right limits at $z=0$. Therefore
continuous solutions of the odd series must satisfy the boundary
condition:
\begin{equation}\label{2.7}
\phi(0)=0.
\end{equation}
Thus, the spectrum of bound states of the odd series coincides
with that of the bound $s$-states of a hydrogen-like atom and this
is also true for the corresponding wave functions on the positive
semi-axis (up to the factor $1/\sqrt{2}$).

However, with the Coulomb singularity the part of Hamiltonian on
the subspace of even functions is not uniquely determined even
under the condition, that the functions from its domain are
continuous everywhere on the real axis, including the point $z=0$.
If the potential in the concerned problem would be nonsingular or,
at least, integrable, then the boundary condition $\phi'(0)=0$
would be natural for the determination of this part. However, any
non-zero at $z=0$ solution of~(\ref{2.6}) is not differentiable at
$z=0$. \footnote{Actually, in one dimension the representation of
Hamiltonian as the sum of operators of kinetic and potential
energies is strictly speaking impossible for potentials with
Coulomb singularities.}

Attempts to choose a proper boundary condition at $z=0$ for the
even part of Hamiltonian by some parity-preserving regularization
of the Coulomb potential in the $\delta$-vicinity of the origin
give non-unique results, depending on ways of regularization and
passing to the limit as $\delta\rightarrow0$. To see this let us
consider the regularized potential
\[
V_\delta(z)=\left\{\begin{array}{c}
-e^2/|z|,\ |z|>\delta;\\
-e^2/|\delta|,\ |z|<\delta.
\end{array}\right.
\]
Since $V_0(z)\leq V_\delta(z)$, then the least eigenvalue
$\mathcal{E}_\mathrm{o}(\delta)$ of the odd series for the
Schr\"{o}dinger operator $\widehat{H}_\delta$ with the potential
$V_\delta$ is not less than the energy of the ground state of the
hydrogen-like atom, that is
\[
-\frac{\mu e^4}{2\hbar^2}\leq\mathcal{E}_\mathrm{o}(\delta).
\]
Let us define further the even part of the Schr\"{o}dinger
operator $\widehat{H}_\delta$ with potential $V_\delta$, assuming
that even functions from its domain satisfy the boundary
condition:
\begin{equation}\label{2.8}
\phi'(-0)+h_\delta\phi(-0)=-\phi'(+0)+h_\delta\phi(+0)=0.
\end{equation}
Taking any
\[
\mathcal{E}_\mathrm{e}<-\frac{\mu e^4}{2\hbar^2}\leq
\mathcal{E}_\mathrm{o}(\delta),
\]
we can arrange by a suitable choice of $h_\delta$ in~(\ref{2.8})
that $\mathcal{E}_\mathrm{e}$ be an eigenvalue of
$\widehat{H}_\delta$. To this end we note, that for $z>\delta$ the
eigenfunction $\phi_\mathrm{e}(z)$, corresponding to the
eigenvalue $\mathcal{E}_\mathrm{e}$, coincides up to a constant
factor with the decreasing as $z\rightarrow\infty$ solution
of~(\ref{2.6}) with $\mathcal{E}$ replaced by
$\mathcal{E}_\mathrm{e}$, that is
\[
\phi_\mathrm{e}(z)=C~W_{k_\mathrm{e},1/2}(2\kappa_\mathrm{e}z),
\]
where $W_{k_\mathrm{e}, 1/2}$ is the Whittaker function. At the
same time for $0<z<\delta$ we have
\begin{eqnarray}
\phi_\mathrm{e}(z)=C'\left[\cos{qz}+\left(h_\delta/q\right)\sin{qz}\right],\nonumber\\
q=\sqrt{\left(2\mu/\hbar^2\right)\left(e^2/\delta-|\mathcal{E}_\mathrm{e}|\right)}\nonumber.
\end{eqnarray}
The continuity condition for the logarithmic derivative of
$\phi_\mathrm{e}(z)$ at $z=\delta$ yields
\begin{equation}\label{2.9}
\begin{split}
h_\delta&=q\left.\frac{2\kappa_\mathrm{e}W'_{k_\mathrm{e},1/2}(2\kappa_\mathrm{e}\delta)\cos{q\delta}+qW_{k_\mathrm{e},1/2}(2\kappa_\mathrm{e}\delta)\sin{q\delta}}{qW_{k_\mathrm{e},1/2}(2\kappa_\mathrm{e}\delta)\cos{q\delta}-2\kappa_\mathrm{e}W'_{k_\mathrm{e},1/2}(2\kappa_\mathrm{e}\delta)\sin{q\delta}}\right|_{\delta\downarrow0}\\
&\approx-2\kappa_\mathrm{e}k_\mathrm{e}\ln(2\kappa_\mathrm{e}k_\mathrm{e}\delta).
\end{split}
\end{equation}
As $\mathcal{E}_\mathrm{e}<\mathcal{E}_\mathrm{o}(\delta)$ and
eigenvalues of even and odd series alternate we conclude that
$\mathcal{E}_\mathrm{e}$ is the least eigenvalue of
$\widehat{H}_\delta$. We see, that by an appropriate choice of
$h_\delta$ we obtain a sequence of Hamiltonians with regularized
potentials and the fixed least eigenvalue.

As a nearest analogue of the boundary condition $\phi'(0)=0$ for
wave functions of the even series we take with account
of~(\ref{2.9}) :
\begin{equation}\label{2.10}
\lim_{z\rightarrow0}\frac{\rmd}{\rmd
z}\left[\left(1+2Az\ln(2Az)\right)\phi(z)\right]=0,
\end{equation}
where $A=e^2\mu/\hbar^2$. The Schr\"{o}dinger differential
operator $\widehat{H}$ in $\mathbf{L}_2(0,\infty)$ defined by this
boundary condition is self-adjoint.

Indeed, using the von Neumann formulae~\cite{akhi} it is easy to
verify, that the functions from the domain
$\mathcal{D}(\widehat{H})$ of each self-adjoint extension
$\widehat{H}$ of the differential operator~(\ref{1.1}) in
$\mathbf{L}_2(0,\infty)$ are continuous on the semi-axis
$[0,\infty)$.

Let $\phi_1,\phi_2\in\mathcal{D}(\widehat{H})$, that is
$\widehat{H}\phi_1,\widehat{H}\phi_2\in\mathbf{L}_2(0,\infty)$. By
continuity of $\phi_1,\phi_2$ at $z=0$ and~(\ref{2.10}) we get
\begin{equation}
\begin{split}
&\int\limits^\infty_0\left[\left(\widehat{H}\varphi_1\right)(z)\varphi^*_2(z)-\varphi_1(z)\left(\widehat{H}\varphi_2\right)^*(z)\right]\rmd
z\\
&=\lim_{\delta\downarrow0}\int\limits^\infty_\delta\left[\left(\widehat{H}\varphi_1\right)(z)\varphi^*_2(z)-\varphi_1(z)\left(\widehat{H}\varphi_2\right)^*(z)\right]\rmd z\\
&=\lim_{\delta\downarrow0}\left[\varphi_1(z)\varphi'^*_2(z)-\varphi'_1(z)\varphi^*_2(z)\right]_{z=\delta}\\
&=\lim_{\delta\downarrow0}\left\{\frac{1}{1+2Az\ln{2Az}}\left[\varphi_1(z)\frac{\rmd}{\rmd z}[(1+2Az\ln{2Az})\varphi^*_2(z)]\right.\right.\\
&\qquad\left.\left.-\varphi^*_2(z)\frac{\rmd}{\rmd
z}[(1+2Az\ln{2Az})\varphi_1(z)]\right]\right\}_{z=\delta}=0\nonumber.
\end{split}
\end{equation}

Therefore $\widehat{H}$ is a symmetric operator. Let us assume
that $\widehat{H}$ is not self-adjoint. Then for each non-real
$\omega$ there is a solution of the equation
\[
-\frac{\hbar^2}{2\mu}\frac{\rmd^2 W}{\rmd
z^2}-\frac{e^2}{|z|}W=\omega W,
\]
which belongs to $\mathbf{L}_2(0,\infty)$ and orthogonal to the
linear set
$\left(\widehat{H}-\omega\right)\mathcal{D}(\widehat{H})$~\cite{akhi}.
But it is easy to verify as above, that such solution is
identically equal to zero. Thus energy levels of the even series
are defined as eigenvalues of Hamiltonian
\[
\widehat{H}=-\frac{\hbar^2}{2\mu}\frac{\rmd^2}{\rmd
z^2}-\frac{e^2}{|z|},
\]
on set of twice differentiable functions $\phi(z)$ at semi-axis
$(0,\infty)$, which satisfy boundary condition~(\ref{2.10}). At
semi-axis $(-\infty,0)$ we, naturally, consider even continuation
of corresponding eigenfunctions.

Evidently, the above choice of the even part of the Hamiltonian is
not exceptional. As it was mentioned above, for example, we can
take the even extension onto the negative semi-axis of the wave
functions satisfying the zero boundary condition at $z=0$ and
obtain in this way an even part of Hamiltonian with the same
spectrum as that for the odd part~\cite{loudon}. The choice of a
concrete boundary condition can be done exceptionally on the base
of physical reasons.

Using the asymptotic expansion of the Whittaker function
$W_{k,1/2}(2\kappa z)$ for $z\rightarrow 0$ we get from
condition~(\ref{2.7}) the eigenvalues of the odd series:
\begin{equation}\label{2.11}
\begin{split}
\frac{1}{\Gamma(1-k)}=0;\
\Rightarrow1-k=-n,\ n=0,1,2 \ldots;\\
\Rightarrow\mathcal{E}_n=-\frac{\mu e^4}{2\hbar^2}\
\frac{1}{n^2},\ n=1,2,3 \ldots,
\end{split}
\end{equation}
and from condition~(\ref{2.10}) the eigenvalues of the even
series:
\begin{equation}\label{2.12}
\mathcal{E}_p=-\frac{\mu e^4}{2\hbar^2} \frac{1}{p^2},
\end{equation}
where $p$, according to~(\ref{2.10}), is defined from equation:
\begin{equation}\label{2.13}
\begin{split}
-p\sum\limits_{j=1}^\infty\frac{1}{j(j-p)}+\frac{1}{2p}-\ln
p+\gamma-1=0,\\
\Rightarrow p=n+1/2+\Delta(n),\ n=0,1,2 \ldots.
\end{split}
\end{equation}
Here $\gamma\simeq0.5772$ is Euler's constant, and $\Delta(n)$ -
is slowly increasing function of the integer number $n$, which
over the range $n\in[0,10]$ obtains values from $-0.013$ to
$0.0156$.

Corresponding normalized wave functions $\phi_n(z)$, which satisfy
the equation~(\ref{2.6}) and the condition~(\ref{2.7}) are given
by:
\begin{equation}\label{2.14}
\begin{split}
&\phi_n(z)=\left(\frac{A}{2n(n-1)^2(n-1)!^2}\right)^{1/2}L_{n-1}^1\left(\frac{2Az}{n}\right)\frac{2Az}{n}\exp\left(-\frac{Az}{n}\right)\qquad n=2,3,4\ldots\\
&\phi_1(z)=\sqrt2 A^{3/2}z\exp(-Az),
\end{split}
\end{equation}
where $L_{n-1}^1\left(2Az/n\right)$ is the generalized Laguerre
polynomial. For the even series, according to~(\ref{2.6})
and~(\ref{2.10}), we obtain:
\begin{equation}\label{2.15}
\phi_p(z)=C_p W_{p, 1/2}\left(\frac{2Az}{p}\right),
\end{equation}
where $C_p$ is normalization factor.

The analytic simplification of~(\ref{2.4}), which depends on the
tube radius $R_0$ but independent of its chirality is the
potential
\begin{equation}\label{2.16}
V_{R_0}(z)=-\frac{e^2}{4\pi^2|z|}\int\limits_0^{2\pi}\int\limits_0^{2\pi}
\frac{\rmd\alpha_1\rmd\alpha_2}{\left(1+(4R_0^2/z^2)\sin^2{\frac{\alpha_1-\alpha_2}{2}}\right)^{1/2}},
\end{equation}
that was obtained from~(\ref{2.4}) under the assumption that the
charges of electron and hole participating in the formation of
exciton are smeared uniformly along infinitesimal narrow bands on
the tube wall. This potential is the simplest approximation to the
bare Coulomb potential, which accounts the finiteness of the tube
diameter. Note, that contrary to the bare Coulomb potential this
one has only logarithmic singularity at the origin. Since
$V_{R_0}(z)$ is an integrable function then solutions of the
Schr\"{o}dinger equation with this potential are continuously
differentiable at $z=0$, and the boundary condition for the even
series in this case is $\phi'(0)=0$. For nanotubes with rather
small diameters the negative eigenvalues of equation~(\ref{2.5})
with potential $V_{R_0}(z)$ appeared to be close to those for
equation~(\ref{2.6}) (see Section 5, table~1 and table~2). For the
both of equations the minimal eigenvalue of the even series well
exceeds the energy gap. This may mean that the single-electron
states in semiconducting SWCNTs in the vicinity of the energy gap
are unstable with regard to the formation of excitons. However,
the screening of $e$-$h$ interaction by the tube electrons could
result in the shift of exciton levels into the gap. To make clear
whether it is so we consider further different forms of screening
of the potential~(\ref{2.16}).

\section{Nanotube dielectric function}
\setcounter{equation}{0}

First we obtain the nanotube dielectric function within the
framework of the Lindhard method (the so-called RPA), then in the
limiting case of small wavenumber values we get the Thomas-Fermi
screening theory for charged particles in semiconducting SWCNTs.

Following the Lindhard method, to obtain the $e$-$h$ interaction
potential $\varphi(\bi{r})$, screened by the electrons of
quasione-dimensional nanotube lattice, we consider the
one-dimensional Fourier transform of the Poisson equation:
\begin{equation}\label{3.1}
\left(q^2-\Delta_\mathrm{2D}\right)\varphi(q,\bi{r}_\mathrm{2D})=4\pi\left(\rho^\mathrm{ext}(q,\bi{r}_\mathrm{2D})+\rho^\mathrm{ind}(q,\bi{r}_\mathrm{2D})\right),
\end{equation}
where $\bi{r}_\mathrm{2D}$ is the transverse component of the
radius-vector, $q$ is the longitudinal component of wave vector,
$\rho^\mathrm{ext}(q,\bi{r}_\mathrm{2D})$ is the one-dimensional
Fourier transform along the tube axis of the density of extraneous
charge $\rho^\mathrm{ext}(z,\bi{r}_\mathrm{2D})$ and
$\rho^\mathrm{ind}(q,\bi{r}_\mathrm{2D})$ is that of the charge
density induced by the extraneous charge. Further we will assume
that $\rho^\mathrm{ext}$ is axial symmetric,
$\rho^\mathrm{ext}(q,\bi{r}_\mathrm{2D})=\rho^\mathrm{ext}(q,r_\mathrm{2D})$,
and localized in the small vicinity of the tube wall. As follows
$\varphi(q,\bi{r}_\mathrm{2D})$ and
$\rho^\mathrm{ind}(q,\bi{r}_\mathrm{2D})$ depend on
$\bi{r}_\mathrm{2D}$ only through $r_\mathrm{2D}$ and besides
whatever the case $\rho^\mathrm{ind}$ is localized at the tube
wall. By~(\ref{3.1}) the screened $e$-$h$ interaction potential
may be written as:
\begin{equation}\label{3.2}
\varphi(q,\bi{r}_\mathrm{2D})=4\pi\int\limits_\mathrm{E_2}\left(\rho^\mathrm{ext}(q,\bi{r}'_\mathrm{2D})+\rho^\mathrm{ind}(q,\bi{r}'_\mathrm{2D})\right)G_0(q,\bi{r}_\mathrm{2D},\bi{r}'_\mathrm{2D})\rmd\bi{r}'_\mathrm{2D},
\end{equation}
where
$G_0(q,\bi{r}_\mathrm{2D},\bi{r}'_\mathrm{2D})=(1/2\pi)K_0(|q||\bi{r}_\mathrm{2D}-\bi{r}'_\mathrm{2D}|)$
is the Green function of the 2D Helmholtz equation, and $K_0$ is
the modified Bessel function of the second kind.

Let $E^0_s(k)$ and $\Psi^0_{k,s}(\bi{r})=(1/\sqrt{N})\exp(\rmi
kz)u^0_{k,s}(\bi{r})$ be the band energies and corresponding Bloch
wave functions of the nanotube $\pi$-electrons and $E_s(k)$,
$\Psi_{k,s}(\bi{r})$ be those in the presence of the extraneous
charge. Then
\begin{equation}\label{3.3}
\begin{split}
\rho^\mathrm{ind}(q,\bi{r}_\mathrm{2D})=-e\int\limits^{L}_{0}&\exp(-\rmi
qz)\sum\limits_{k,s}\left[f(E_s(k))|\Psi_{k,s}(\bi{r})|^2\right.\\
&\left.-f(E^0_s(k))|\Psi^0_{k,s}(\bi{r})|^2\right]\rmd z,
\end{split}
\end{equation}
where $f$ is the Fermi-Dirac function, $L$ is the length of CNT
and $s$ numbers single-electron bands ($N$ is the number of unit
cells in the nanotube). In the linear in $\varphi$ approximation
we get:
\begin{equation}\label{3.4}
\begin{split}
\rho^\mathrm{ind}(q,\bi{r}_\mathrm{2D})=-\frac{e^2}{L}\sum\limits_{k,s,s'}&\frac{B_{s,s'}(k,k-q,a)}{E_{\mathrm{g};s,s'}(k)}\\
&\times\int\limits^{a}_{0}u^*_{\mathrm{v};k-q,s}(z,\bi{r}_\mathrm{2D})u_{\mathrm{c};k,s'}(z,\bi{r}_\mathrm{2D})\rmd
z\varphi(q,R_0),
\end{split}
\end{equation}
where $a$ is the longitudinal period of nanotube and
\begin{equation}
\begin{split}
&B_{s,s'}(k,k-q,a)=\int\limits_\mathrm{E_2}\int\limits^{a}_{0}u^*_{\mathrm{c};k,s'}(z,\bi{r}_\mathrm{2D})u_{\mathrm{v};k-q,s}(z,\bi{r}_\mathrm{2D})\rmd
z\rmd\bi{r}_\mathrm{2D},\\
&E_{\mathrm{g};s,s'}(k)=E_{\mathrm{c};s'}(k)-E_{\mathrm{v};s}(k).\nonumber
\end{split}
\end{equation}
Taking into account the axial symmetry of
$\rho^\mathrm{ext}(q,\bi{r}_\mathrm{2D})$ and
$\rho^\mathrm{ind}(q,\bi{r}_\mathrm{2D})$ and their localization
near the nanotube wall $(r_\mathrm{2D}=R_0)$ we obtain
from~(\ref{3.2}) and~(\ref{3.4}) that
\begin{equation}\label{3.5}
\varphi(q,R_0)=\widetilde{\varphi}(q,R_0)+2I_0(|q|R_0)K_0(|q|R_0)\int\limits_\mathrm{E_2}\rho^\mathrm{ind}(q,\bi{r}_\mathrm{2D})\rmd\bi{r}_\mathrm{2D},
\end{equation}
where $\widetilde{\varphi}(q,R_0)$ is the Fourier transform of the
electrostatic potential induced by $\rho^\mathrm{ext}$, and $I_0$
is the modified Bessel function of the first kind. We see that:
\begin{equation}\label{3.6}
\begin{split}
&\varphi(q,R_0)=\frac{\widetilde{\varphi}(q,R_0)}{\varepsilon_{R_0,a}(q)},\\
&\varepsilon_{R_0,a}(q)=1+\frac{e^2}{\pi}\sum\limits_{s,s'}\int\limits^{\pi/a}_{-\pi/a}\frac{|B_{s,s'}(k,k-q,a)|^2}{E_{\mathrm{g};s,s'}(k)}\rmd
k\ I_0(|q|R_0)K_0(|q|R_0).
\end{split}
\end{equation}
In the limiting case of small wavenumbers:
\begin{equation}\label{3.7}
\begin{split}
&|B_{s,s'}(k,k-q,a)|^2_{q\rightarrow0}\approx\left|U_{s,s'}(k,a)\right|^2\ q^2,\\
&\left|U_{s,s'}(k,a)\right|^2=\left|\
\int\limits_\mathrm{E_2}\int\limits^{a}_{0}u^*_{\mathrm{c};k,s'}(z,\bi{r}_\mathrm{2D})\frac{\partial}{\partial
k}u_{\mathrm{v};k,s}(z,\bi{r}_\mathrm{2D})\rmd
z\rmd\bi{r}_\mathrm{2D}\ \right|^2.
\end{split}
\end{equation}
Note that $U_{s,s'}(k,a)$ is nonzero only for the mirror bands,
that is $U_{s,s'}(k,a)=U_s(k,a)\delta_{s,s'}$. Using the
orthogonality of the Bloch wave functions and applying the
Schr\"{o}dinger equation for $\Psi_{k,s}(z,\bi{r}_\mathrm{2D})$
yields
\begin{equation}\label{3.8}
U_s(k,a)=\frac{\rmi\hbar^2N}{m_\mathrm{e}
E_{\mathrm{g};s,s}(k)}\int\limits_\mathrm{E_2}\int\limits^{a}_{0}\Psi^*_{\mathrm{c};k,s}(z,\bi{r}_\mathrm{2D})\frac{\partial}{\partial
z}\Psi_{\mathrm{v};k,s}(z,\bi{r}_\mathrm{2D})\ \rmd
z\rmd\bi{r}_\mathrm{2D}.
\end{equation}
Hence, the screened quasione-dimensional electrostatic potential
induced by a charge $e_0$, distributed with the density:
\[
\rho^\mathrm{ext}(\bi{r})=\frac{e_0}{2\pi
R_0}\delta(z)\delta(r_\mathrm{2D}-R_0),\nonumber
\]
in accordance with~(\ref{3.6}) and~(\ref{3.7}), is given by the
expression
\begin{equation}\label{3.9}
\varphi(z)=\frac{e_0}{\pi
R_0}\int\limits^{\infty}_{-\infty}\frac{I_0(|q|)K_0(|q|)\exp(\rmi
qz/R_0)}{1+g_aq^2I_0(|q|)K_0(|q|)}\ \rmd q,
\end{equation}
with
\begin{equation}\label{3.10}
g_a=\frac{e^2\hbar^4}{\pi m^2_\mathrm{e}
R^2_0}\sum\limits_s\int\limits^{\pi/a}_{-\pi/a}\frac{1}{E^3_{\mathrm{g};s,s}(k)}
\left|\left\langle\Psi_{\mathrm{c};k,s}\left|\frac{\partial}{\partial
z}\right|\Psi_{\mathrm{v};k,s}\right\rangle\right|^2\rmd k.
\end{equation}
This potential was calculated using the single-electron energy
spectrum and wave functions, obtained in~\cite{tish}. The ground
state exciton binding energy, calculated from~(\ref{2.5}) with the
screened potential~(\ref{3.9}), remains noticeably greater than
the energy gap (see Section 5, table~4).

\section{Screening by free charges}
\setcounter{equation}{0}

Free charges may appear in semiconducting nanotubes at rather high
temperatures $T$. So here we will obtain the self-consistent
screened potential of $e$-$h$ interaction depending on the
nanotube diameter and medium temperature.

To take into account the screening of $e$-$h$ interaction
potential $\varphi(\bi{r})$ by free charges (by intrinsic
electrons and holes) we consider the Poisson equation:
\begin{equation}\label{4.1}
-\Delta\varphi+\kappa^2 R_0\delta(r-R_0)\varphi=4\pi
e\delta(\bi{r}-\bi{r}_0),
\end{equation}
where we suppose again, that the screening particles (electrons
and holes) and the screened $e$-$h$ pair itself are localized at
the surface of cylinder (nanotube's wall) with radius $R_0$. Here
$\kappa^2=(4\pi e^2 n_0/k_\mathrm{B}T)(1/\pi R^2_0)$, and
\[
n_0=\left(\sqrt{2\pi\sqrt{m^*_\mathrm{h}
m^*_\mathrm{e}}k_\mathrm{B} T}/2\pi
\hbar\right)\exp(-E_\mathrm{g}/2k_\mathrm{B} T)\nonumber
\]
is the one-dimensional analogue of particle concentration in the
intrinsic semiconductors. We assume, that CNTs can be treated as
such semiconductors.

Equation~(\ref{4.1}) (without factor $4\pi e$) can be represented
in the equivalent form:
\begin{equation}\label{4.2}
G(\bi{r},\bi{r}_0)=G_0(\bi{r},\bi{r}_0)-\kappa^2 R_0
\int\limits_\mathrm{E_3}G_0(\bi{r},\bi{r}')\delta(r'-R_0)G(\bi{r}',\bi{r}_0)\rmd\bi{r}',
\end{equation}
where $G_0(\bi{r},\bi{r}_0)=1/4\pi|\bi{r}-\bi{r}_0|$ is the Green
function of the Poisson equation without screening $(\kappa=0)$.
After averaging over axial and radial components of the
radius-vector and several Fourier transforms, we obtain the
following one-dimensional screened $e$-$h$ interaction potential
$\varphi(z)$:
\begin{equation}\label{4.3}
\varphi(z)=\frac{1}{\sqrt{2\pi}}\int\limits^{\infty}_{-\infty}\frac{\widetilde{\varphi}_0(k)\exp(\rmi
kz)}{1+(2\pi)^{3/2}(\kappa R_0)^2 \widetilde{\varphi}_0(k)/4\pi
e}\ \rmd k,
\end{equation}
where $\widetilde{\varphi}_0(k)$ is the Fourier transform of the
average unscreened potential~(\ref{2.16}):
\begin{equation}\label{4.4}
\begin{split}
\widetilde{\varphi}_0(k)&=\frac{4\pi
e}{8\pi^2}\int\limits^{2\pi}_{0}\frac{1}{\sqrt{2\pi}}\int\limits^{\infty}_{-\infty}\frac{\exp(-\rmi
k\tilde{z})}{|\tilde{z}|\left(1+(4R^2_0/\tilde{z}^2)\sin^2(\alpha/2)\right)^{1/2}}\ \rmd\tilde{z}\rmd\alpha\\
&=\frac{4\pi e}{(2\pi)^{3/2}}I_0(|k|R_0)K_0(|k|R_0),
\end{split}
\end{equation}
where $I_0(|k|R_0)$ and $K_0(|k|R_0)$ are the same modified Bessel
functions of the first and the second kind respectively. Hence,
the $e$-$h$ interaction potential screened by free charges for any
semiconducting SWCNT is given by:
\begin{equation}\label{4.5}
\varphi(z)=\frac{e}{\pi R_0}
\int\limits^{\infty}_{-\infty}\frac{I_0(|k|)K_0(|k|)\exp(\rmi
kz/R_0)}{1+(\kappa R_0)^2 I_0(|k|)K_0(|k|)}\ \rmd k.
\end{equation}
The screened potential~(\ref{4.5}) can be used for the calculation
of the large radius exciton binding energies in the ground and
excited states for the large-diameter SWCNTs at high temperatures.
The ground state binding energy, calculated from~(\ref{2.5}) with
the screened potential~(\ref{4.5}), remains greater than the
energy gap (see Section 5, table~5).

To compare different obtained potentials, we produce figure~1,
that shows them plotted point by point for the semiconducting
(28,0) nanotube in comparison with the bare Coulomb potential.
Figure~1 also shows that the mentioned above screened potentials
slightly differ from the bare Coulomb potential when the distance
between electron and hole is large. This fact justifies the bare
Coulomb large radius exciton model given in the beginning.
\begin{figure}[t]
\begin{center}
\includegraphics[scale=1.25]{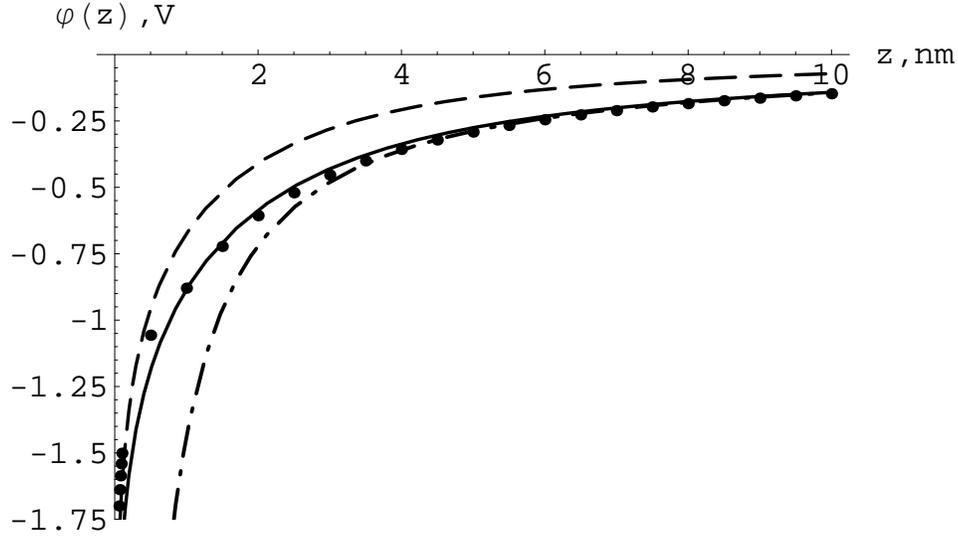}
\caption{\label{fig1} The $e$-$h$ interaction potentials versus
the electron-hole distance $z$ for the zig-zag nanotube (28,0):
dashed line - the screened one calculated by~(\ref{4.5}) for
$T=550^\circ\mathrm{K}$; black circles - the screened
potential~(\ref{3.9}); solid line - the Coulomb unscreened
averaged potential~(\ref{2.16}); dot-dashed line - the bare
Coulomb potential from~(\ref{2.6})}
\end{center}
\end{figure}

\section{Calculation results. Screening influence}
\setcounter{equation}{0}

Electronic structure of nanotubes, electron and hole effective
masses and energy gap magnitudes were obtained in~\cite{tish}
within the framework of the zero-range potential method for the
Bloch wave functions~\cite{aghh}. Using those values of effective
masses and energy gaps, we have calculated the unscreened and
screened $e$-$h$ interaction potentials and corresponding exciton
binding energies of the ground and excited states, which either
explicitly or implicitly depend on parameters of concrete
semiconducting SWCNT (chirality, radius, reduced effective mass,
band gap magnitude) and the temperature of medium (in Section 4).
Here we present results of these calculations.

Numerically calculated values of the exciton binding energies of
the ground and excited states according to~(\ref{2.6}) are given
in table 1.

\begin{table}[h]
\caption{\label{tab1} Exciton binding energies according
to~(\ref{2.6}).}
\begin{center}
\begin{tabular}{|c|c|c|c|c|c|}
\hline
$\mathrm{Chirality}$&$E_\mathrm{g},~\mathrm{eV}$&$\mathcal{E}_{0;\mathrm{even}},~\mathrm{eV}$&$\mathcal{E}_{1;\mathrm{odd}},~\mathrm{eV}$&$\mathcal{E}_{1;\mathrm{even}},~\mathrm{eV}$&$\mathcal{E}_{2;\mathrm{odd}},~\mathrm{eV}$\\
\hline
(7,0)&1.3416&-2.8343&-0.6722&-0.294&-0.168\\
\hline
(6,5)&1.1017&-2.9253&-0.6938&-0.3034&-0.1734\\
\hline
(28,0)&0.3674&-0.9484&-0.2249&-0.0983&-0.0562\\
\hline
\end{tabular}
\end{center}
\end{table}

These results unambiguously show that the binding energies in the
even ground state for any of the selected semiconducting SWCNTs
are much greater than the corresponding energy gaps in the bare
Coulomb limit~(\ref{2.6}).

Further, the numerically calculated values of exciton binding
energies at the ground and excited states according to the wave
equation with the potential~(\ref{2.16}) are given in table~2. It
can be seen from table~2, that the discrepancies with the
analogous results in table~1 are more considerable for nanotubes
with larger diameters, because the wave equation with the
potential~(\ref{2.16}) tends to~(\ref{2.6}) if $R_0\rightarrow0$.

\begin{table}[h]
\caption{\label{tab2} Exciton binding energies according to the
Schr\"{o}dinger equation with the potential~(\ref{2.16})}
\begin{center}
\begin{tabular}{|c|c|c|c|c|c|}
\hline
$\mathrm{Chirality}$&$2R_0,~\mathrm{nm}$&$\mathcal{E}_{0;\mathrm{even}},~\mathrm{eV}$ &  $\mathcal{E}_{1;\mathrm{odd}},~\mathrm{eV}$ & $\mathcal{E}_{1;\mathrm{even}},~\mathrm{eV}$& $\mathcal{E}_{2;\mathrm{odd}},~\mathrm{eV}$\\
\hline
(7,0)&0.548&-2.7894&-0.5365&-0.2955&-0.1507\\
\hline
(6,5)&0.7468&-2.2503&-0.5071&-0.2829&-0.1488\\
\hline
(28,0)&2.192&-0.7567&-0.1668&-0.0928&-0.0486\\
\hline
\end{tabular}
\end{center}
\end{table}

We can see also from table~2 that the ground state binding
energies are larger than the corresponding energy gaps even if the
finiteness of nanotubes is taken into account.

The next table (table~3) shows that the exciton radii are
comparable with the corresponding nanotubes diameters, thus they
much greater than the nanotube lattice parameter 0.142~nm.
Therefore the large radius exciton theory methods are appropriate
for the treatment of the SWCNTs exciton problem.

\begin{table}[h]
\caption{\label{tab3} Exciton radii $r_n\sim n/2A$ in units of
$2R_0$}
\begin{center}
\begin{tabular}{|c|c|c|c|c|}
\hline
$\mathrm{Chirality}$&$r_{0;\mathrm{even}}$&$r_{1;\mathrm{odd}}$&$r_{1;\mathrm{even}}$ & $r_{2;\mathrm{odd}}$\\
\hline
(7,0)&0.4759&0.9772&1.4776&1.9545\\
\hline
(6,5)&0.3383&0.6948&1.05&1.3895\\
\hline
(28,0)&0.3556&0.73&1.1&1.46\\
\hline
\end{tabular}
\end{center}
\end{table}

As illustration we have calculated the binding energies for the
(28,0) zig-zag nanotube with account of the nanotube dielectric
function (table~4).

\begin{table}[h]
\caption{\label{tab4}Exciton binding energies for nanotube (28,0)
according to~(\ref{3.9}) and~(\ref{3.10})}
\begin{center}
\begin{tabular}{|c|c|c|c|c|c|}
\hline
$\mathrm{Chirality}$ & $g_a$ & $\mathcal{E}_{0;\mathrm{even}},~\mathrm{eV}$ &  $\mathcal{E}_{1;\mathrm{odd}},~\mathrm{eV}$ &$ \mathcal{E}_{1;\mathrm{even}},~\mathrm{eV}$& $\mathcal{E}_{2;\mathrm{odd}},~\mathrm{eV}$\\
\hline
(28,0)& 0.6 & -0.6869 & -0.1799 & -0.0952 & -0.0501\\
\hline
\end{tabular}
\end{center}
\end{table}

The data from table~4 obviously show that the screening by
nanotube band electrons is not enough for the ground state exciton
binding energy to be less than the energy gap.

The exciton binding energies at the ground and excited states for
the semiconducting (28,0) SWCNT calculated using the
potential~(\ref{4.5}) for $T=550^\circ\mathrm{K}$ are listed in
table~5.

\begin{table}[h]
\caption{\label{tab5}Exciton binding energies for the nanotube
(28,0) according to~(\ref{4.5}) for $T=550^\circ\mathrm{K}$}
\begin{center}
\begin{tabular}{|c|c|c|c|c|c|}
\hline
$\mathrm{Chirality}$ & $\kappa R_0$ & $\mathcal{E}_{0;\mathrm{even}},~\mathrm{eV}$ &  $\mathcal{E}_{1;\mathrm{odd}},~\mathrm{eV}$ &$ \mathcal{E}_{1;\mathrm{even}},~\mathrm{eV}$& $\mathcal{E}_{2;\mathrm{odd}},~\mathrm{eV}$\\
\hline
(28,0)& 0.38 & -0.5549 & -0.0642 & -0.0294 & -0.0133\\
\hline
\end{tabular}
\end{center}
\end{table}

Note, that the screened potential~(\ref{4.5}) may be used either
for the semiconducting SWCNTs with narrow band gap (as zig-zag
$(3n,0)$ SWCNTs) or for the large-diameter nanotubes (small gaps)
or(and) at rather high temperatures, because only under these
conditions the linear concentration $n_0$ of free charged
particles provides a perceptible screening. At
$T=550^\circ\mathrm{K}$ the (28,0) nanotube has approximately one
free charged particle per micrometer of its length, but even at
these conditions the screening by free charges of the $e$-$h$
interaction potential is much stronger than the screening by the
all bound electrons of semiconducting SWCNTs (compare table~4 and
table~5). Nevertheless, as it follows from the same table~5, even
in this case the ground state exciton binding energy still exceeds
the energy gap.

\section{Discussion}
\setcounter{equation}{0}

In the all above examples the binding energy of the ground state
of even excitons in isolated SWCNTs appeared to be much greater
than the corresponding band gaps even with account of some
screening effects by tubes $\pi$-electrons. This should mean that
the single-electron states in SWCNTs are unstable at least in the
vicinity of the energy gap with respect to formation of excitons.
Such conclusion might seem doubtful though we came to it applying
similar arguments as in the case of 3D large radius excitons.
There are three reasons due to which a partial destruction of band
electrons states in semiconducting SWCNTs  in reality is either
absent or inconspicuous.

First of all the account of dynamical screening, that is the
frequency dependence of dielectric function, may return the all
exciton levels into the band gap. This was shown in~\cite{spb},
where calculations of the exciton binding energy with the static
dielectric function yielded also the exciton binding energy
exceeding the energy gap. At the same time the self-consistent
calculation with frequency dependent dielectric function gave
according to~\cite{spb} a universal ratio of the exciton binding
energy to the energy gap depending only on the resonance integral
$\gamma_0$ but not on the nanotube radius (it equals 0.87 if
$\gamma_0=2.7$~eV). By~\cite{spb} the exciton binding energy
cannot be larger than the energy gap because of the singularity of
the frequency-dependent dielectric function $\varepsilon(\omega)$
at $\omega=E_\mathrm{g}/\hbar$ for the frequencies, corresponding
to the direct transitions between the van Hove points of the tube
single-electron spectral density. However, actually this argument
is true only if the exciton binding energy obtained without
account of dynamical screening gets into a small vicinity of the
energy of allowed transition between such points. This is because
the frequency dependent SWCNT dielectric function may only then
become rather great. Otherwise as it follows from results
of~\cite{atish} the effect of dynamical screening is too small and
the exciton state with the binding energy much greater than the
energy gap transforms into a long-living resonance in the
continuous spectrum of electron-hole pairs with opposite
quasi-momenta.

The second reason is the so-called environmental effect. In
experimental works~\cite{bachilo}-\cite{wang} (which used the
methods described in~\cite{fluor}) investigated individual
nanotubes were not in vacuum but encased in sodium dodecyl sulfate
(SDS) cylindrical micelles disposed in $\mathrm{D_2O}$. Because of
these SDS micelles, which provided a pure hydrocarbon environment
around individual nanotubes, the high permittivity solvent
$\mathrm{D_2O}$ did not reach nanotubes. However, the environment
of hydrophobic hydrocarbon "tales" $(-\mathrm{C_{12}H_{25}})$ of
the SDS molecules has the permittivity greater than unity.
Following the figure~1A from~\cite{fluor} we considered a simple
model of a SWCNT in a dielectric environment: a hollow, narrow,
infinite cylinder with radius $R_0$ in a medium with the
dielectric constant $\varepsilon$ and found the
potential~(\ref{2.16}) screened by the medium within the framework
of mentioned model under the assumption about axially symmetrical
charge localization at nanotube's (here - cylinder's) wall. The
corresponding 1D screened potential $\varphi(z)$ is given by:
\begin{equation}\label{6.1}
\varphi(z)=-\frac{e}{\pi R_0}
\int\limits^{\infty}_{-\infty}\frac{I_0(|k|)K_0(|k|)\exp(\rmi
kz/R_0)}{\left[\varepsilon
K_1(|k|)I_0(|k|)+I_1(|k|)K_0(|k|)\right]|k|}\ \rmd k,
\end{equation}
where $I_j(|k|)$ and $K_j(|k|)$ are the modified Bessel functions
of the order $j$ of the first and the second kind, respectively.
We don't know the exact value of dielectric constant of the pure
medium, which is formed from the hydrocarbon "tales" of the SDS
molecules. But for estimates we take the dielectric constants of
the substances, which are also formed from similar hydrocarbon
"tales", e.g.: petroleum $(\varepsilon\simeq2.1)$ or dodecane
$(\varepsilon\simeq2)$ at $293^\circ\mathrm{K}$ (this temperature
is very close to that used in~\cite{fluor}-\cite{bachilo2}), or
polyethylene $(\varepsilon\simeq2.2-2.4)$. Using the
potentials~(\ref{2.16}) and~(\ref{6.1}) with $\varepsilon$ varying
in the interval $2-2.4$ we have got that the ground state exciton
binding energy in the nanotube (8,0) (the energy gap equals
1.415~eV~\cite{tish}) is 3.06~eV in vacuum while with account of
the environment it runs the interval $1.33-1.06$~eV and hence gets
into the corresponding energy gap and  becomes close to those
in~\cite{spataru2} (about $0.86-1$~eV), even without account of
static and dynamical dielectric screening of the
potential~(\ref{2.16}) by nanotube electrons. Remind that results
on the (8,0) nanotube in~\cite{spataru2} are in good agreement
with those obtained  in~\cite{bachilo2} by interpolation of
experimental data for another species of nanotubes.

Further, taking the (7,5) nanotube we compare our results with the
corresponding experimental data from~\cite{wang}, where individual
SWCNTs were isolated in surfactant micelles of SDS in
$\mathrm{D_2O}$ like in~\cite{fluor}. Our calculations for the
(7,5) nanotube in vacuum yield 2.12~eV as the ground state exciton
binding energy, while for the same tube in the SDS environment the
binding energy calculated using the potential~(\ref{6.1}) gets
into the interval $0.90-0.71$~eV (the band gap for the (7,5) tube
is 1.01~eV~\cite{tish}) depending on $\varepsilon$ varying from 2
to 2.4. The obtained binding energy value is not far from that
of~\cite{wang} $\sim0.62$~eV even without the account of static
and dynamical dielectric screening of the potential~(\ref{2.16})
by the nanotube electrons. There is a comparison of experimental
data on the exciton binding energies in the work~\cite{wang} with
the corresponding theoretical results of~\cite{zhao}. These
results are well agreed. But again, in the work~\cite{zhao} the
interparticle potential includes screening parameter denoted as
$\kappa=2$. Besides, it is asserted in~\cite{zhao} that the
assumption of similar Coulomb parameters for SWCNTs and
phenyl-based $\pi$-conjugated polymers, used in this work, gives
smaller exciton binding energies for SWCNTs. All the results
listed in table~1 - table~5 of our work are related only to SWCNTs
in vacuum. So let us turn to the experimental work~\cite{ohno}
which deals with optical properties (photoluminescence) of SWCNTs
suspended in air (near-unit dielectric constant). As it follows
from~\cite{ohno}, the relative discrepancies between the optical
transition energies obtained in~\cite{ohno} and those obtained
in~\cite{bachilo2} are not significant (about several percents).
This result could be expected, since according to the usual
self-consistent field approximations the interaction of a
$\pi$-electron with other electrons of a nanotube should be
substantially compensated in the ground state by the interaction
with the nearest ions. Evidently, the effect of this compensation
is not sensitive to an environmental screening. However, for
excited states such as excitons, where electrons and holes are at
distances of the order of tube diameter, the environmental effect
can be strong.

Note thirdly that with the advent of $N$ excitons in the tube the
additional screening effect, stipulated by a rather great
polarizability of excitons in the longitudinal electric field,
appears. The elementary estimates show that the corresponding
adding to the dielectric constant is
\[
\Delta\varepsilon\approx4\pi\frac{Ne^2}{\mathcal{E}_\mathrm{b}L},
\]
where $\mathcal{E}_\mathrm{b}$ is the binding energy of even
exciton in the ground state and $L$ is the length of a tube. We
see that in the case of $N\sim10$ per 100~nm of nanotube length
$\Delta\varepsilon\simeq1$ and therefore the lowest exciton
binding energy occurs already inside the energy gap. This blocks
further conversions of single-electron states into excitons. The
shift of the forbidden band edges due to the transformation of
some single-electron states into excitons results in some
enhancement of the energy gap. As follows the optical transition
energy $E_{11}$ should be blueshifted as in~\cite{ohno}. A coarse
estimate of this shift using the elementary relation
\[
\Delta E_{11}\approx\frac{\hbar^2\pi^2N^2}{\mu L^2}
\]
gives $\Delta E_{11}/E_{11}\sim10^{-2}$. If the exciton gas in
tubes is unstable with respect to transition into a
one-dimensional electron-hole plasma, then for the account of
screening effect produced by this plasma we can use the results of
Section 4. For example, for the (8,0) tube even ten charges per
100~nm of its length ($\sim0.1\%$ of $\pi$-electrons number)
reduce the ground state exciton binding energy to $0.12$~eV and
thus block spontaneous transitions to the exciton states.

Thus we may conclude that the ground state of $\pi$-electrons in
semiconducting SWCNTs in vacuum is formed by band electrons
filling all the levels up to a certain level below the gap
together with some amount of two-particle even excitations, which
can form either a rare gas of excitons or electron-hole plasma.
The additional screening effect induced by the exciton gas (or the
one-dimensional $e$-$h$ plasma) blocks further partial destruction
of single-electron states. The environmental effect may return the
even exciton binding energies into the energy gap and thus may
remove two-particle excitations from the ground state of
$\pi$-electrons in SWCNTs.

\section*{Acknowledgements}

The authors would like to thank Sergey Tishchenko for assistance
with some numerical calculations. This work was partly supported
by the Civilian Research and Development Foundation of USA (CRDF)
and the Government of Ukraine, grant UM2-2811-OD06.

\end{document}